# Qutrit State Space Sections using 3-Dimensional Vectors

**Vinod K. Mishra, Ph.D.**

*US Army Research Laboratory, Aberdeen, MD 21005*

*vinod.k.mishra.civ@mail.mil*

## 1. Introduction

The qutrit having three internal states comes next in complexity after qubit as a resource for quantum information processing.

A qutrit state density matrix is of order 3 and depends on 8 parameters. Whereas the qubit density matrix can be easily visualized using Bloch sphere representation of its states, at the same time this simplicity is unavailable for the 8-dimensional Qutrit State Space (QtSS). Still there have been several attempts at visualizing the QtSS [2-7] in the past.

(i) 8 Gell-Mann matrices, which form a complete set for expressing 3 × 3 SU(3) matrices were used [3-4].

(ii) 3-dimensional Bloch matrices and their principal minors [7]

(iii) 6 Gell-Mann matrices supplemented by 2 matrices based on Spin-1 representation replacing two diagonal Gell-Mann ones [5-6].

Attempts were made to capture the complexity of QtSS by studying its 2 and 3 dimensional sections. The sections of the 8-dimensional QtSS are defined as the lower dimensional spaces when the number of non-zero parameters is less than 8. The study of 2- and 3-sections led to many interesting geometric shapes in 2 and 3 dimensions. Unfortunately this approach could not be extended to 4 and higher order sections because of the difficulties in visualizing them.

Recently an alternative approach to study the QtSS with 3 dimensional vectors defined using constraints on the density matrix has become available [1]. Some preliminary results on 2-sections based on them was also presented. In the present work we extend that investigation to recast the SU(3) based well-known 2- and 3-section results in terms of new 3-dimensional vectors, and also present representative analysis of sections of order higher than 3.

## 2. New 3-Dimensional Qutrit State Vectors

The most general Qutrit Density Matrix (QtDM) can be represented using the well-known SU (3) invariant form as $\rho = \frac{1}{3}I_3 + \vec{n}.\vec{\lambda}$. Here $I_3$ is $3 \times 3$ unit matrix, , $\vec{n} = (n_1, n_2, \ldots, n_8)$ is an 8-dimensional vector with 8 real components, and , $\vec{\lambda} = (\lambda_1, \lambda_2, \ldots, \lambda_8)$ is a matrix vector with $3 \times 3$ Gell-Mann matrices as its components. The same QtDM can also be represented alternatively in terms of the symmetric part of the 2 qubit Bloch matrix representing a spin-1 state.

$$\rho = \begin{bmatrix} \omega_1 & (q_3 + ia_3)/2 & (q_2 - ia_2)/2 \\ (q_3 - ia_3)/2 & \omega_2 & -(q_1 + ia_1)/2 \\ (q_2 + ia_2)/2 & -(q_1 - ia_1)/2 & \omega_3 \end{bmatrix}$$

The parameters in this representation are connected to spin-1 observables:
$\omega_i =<S_i^2> = Tr(\rho S_i^2)$, $a_i =<S_i> = Tr(\rho S_i)$, and
$q_k =<S_i S_j + S_j S_i> = Tr\{\rho(S_i S_j + S_j S_i)\}, k \neq i, j$.

Application of constraints on QtDM leads to the following results.

    (i)    Constraint due to 1st Invariant: $I_1 = Tr(\rho) = 1$

This constraint is built into the SU (3) representation through the definition of n3 and n8 matrices. For spin-1 representation, we get $\omega_1 + \omega_2 + \omega_3 = 1$. For a physical QtDM we also have $\omega_1 \geq 0, \omega_2 \geq 0, \omega_3 \geq 0$.

One can capture these constraint properties using a 3-dimensional vector defined as

$$\vec{w} = \{\sqrt{\omega_1}, \sqrt{\omega_2}, \sqrt{\omega_3}\},: Tr(\rho) = \omega_1 + \omega_2 + \omega_3 = w^2 = 1$$

    (ii)    Constraint due to 2nd Invariant; $I_2 = Tr(\rho^2) \leq 1$ (equality sign for pure states)

In terms of Gell-Mann parameters, one finds

$$I_2 = Tr(\rho^2) = \sum_{i=1}^{8} n_i^2 \leq 1$$

Define $r_i^2 = \frac{1}{4}(a_i^2 + q_i^2)$, then the constraint in terms of new parameters takes the following form.

$$I_2 = Tr(\rho^2) = \sum_{i=1}^{3}(\omega_i^2 + 2r_i^2) \leq 1$$

Let us define a vector $\vec{u}$ such that $\vec{u} = \{u_1, u_2, u_3\}$ with

$$u_1 = \sqrt{\frac{1}{3} - 2(\omega_2 \omega_3 - r_1^2)}, \quad u_2 = \sqrt{\frac{1}{3} - 2(\omega_3 \omega_1 - r_2^2)}, \quad u_3 = \sqrt{\frac{1}{3} - 2(\omega_1 \omega_2 - r_3^2)}$$

We have used the following relation in the expressions above.

$$\sum_{i=1}^{3} \omega_i^2 = 1 - 2(\omega_2\omega_3 + \omega_3\omega_1 + \omega_1\omega_2)$$

Then $I_2 = Tr(\rho^2) = u^2 = \sum_{i=1}^{3}(\omega_i^2 + 2r_i^2)$.

(iii) Constraint due to 3rd Invariant: $I_3 = 3Tr(\rho^2) - 2Tr(\rho^3) \leq 1$

The equality sign in the above constraint corresponds to the boundary surface of the 8-dimensional QtSS. In the Gell-Mann representation it is given as

$$I_3 = 3Tr(\rho^2) - 2Tr(\rho^3) = 3\sum_{i=1}^{8} n_i^2 - 6n_8\left(\sum_{i=1}^{3} n_i^2 - \frac{1}{2}\sum_{i=4}^{7} n_i^2 - \frac{1}{3}n_8^2\right)$$

$$-6\sqrt{3}(n_1 n_4 n_6 + n_1 n_5 n_7 + n_2 n_5 n_6 - n_2 n_4 n_7) - 3\sqrt{3}n_3(n_4^2 + n_5^2 - n_6^2 - n_7^2) = 1$$

Define

$$D \equiv a_2 a_3 q_1 + a_3 a_1 q_2 + a_1 a_2 q_3 - q_1 q_2 q_3.$$

Then $I_3$ can be recast in terms of new variables as

$$I_3 = 3Tr(\rho^2) - 2Tr(\rho^3) = \sum_{i=1}^{3}(3\omega_i^2 - 2\omega_i^3 + 6\omega_i r_i^2) - \frac{3}{2}D = 1,$$

The above expression can be rewritten in terms of a 3-dimensional vector $\vec{v} = \{v_1, v_2, v_3\}$. Define

$$X \equiv 1/3 - 1/2 (4\omega_1\omega_2\omega_3 + D)$$

The following relation has been used above.

$$\sum_{i=1}^{3}(3\omega_i^2 - 2\omega_i^3) = 1 - 6\omega_1\omega_2\omega_3$$

Then the components of $\vec{v}$ are given as

$$v_1 = \sqrt{X + 6\omega_1 r_1^2}, \; v_2 = \sqrt{X + 6\omega_2 r_2^2}, \; v_3 = \sqrt{X + 6\omega_3 r_3^2}, \text{ with } v^2 = 1 \text{ (QtSS boundary)}.$$

*Relation of Invariants to the 3-Vectors*

The relations are given as

$$u^2 = \frac{1}{3} + \frac{2}{3}I_2 =, v^2 = \frac{7}{9} + \frac{2}{9}I_3$$

*Relation of Density Matrix to the Constraints*

There are 3 situations for a given density matrix $\rho$ in relation to these constraints.

(i) The given $\rho$ is a density operator if and only if

$$Tr(\rho) = 1, Tr(\rho^2) \leq 1, 3Tr(\rho^2) - 2Tr(\rho^3) \leq 1$$

(ii) The given $\rho$ is a density operator for a boundary state if and only if

$$Tr(\rho) = 1, Tr(\rho^2) \leq 1, 3Tr(\rho^2) - 2Tr(\rho^3) = 1$$

(iii) The given $\rho$ is a density operator for a pure state if and only if

$$Tr(\rho) = 1, Tr(\rho^2) = 1, 3Tr(\rho^2) - 2Tr(\rho^3) = 1$$

The states satisfying the $I_3$ constraint have to be checked against the $I_2$ constraint to decide if they are pure. In the original 8-dimensions it means that the intersection of the geometric objects due to the $I_2$ and $I_3$ constraints contains the pure state space.

These vectors are always in the positive octant of a 3-dimensional sphere of unit radius. The mixed states reside inside the volume interior to the "pure qutrit state surface". This is similar to the Bloch ball of a qubit.

### *Angular Representation for Pure States*

The property $w^2 = 1$ can be satisfied by an angular representation for the vector $\vec{w}$ as.

$\vec{w} = \{sin\theta cos\phi, sin\theta sin\phi, cos\theta\}$, so that

$\omega_1 = sin^2\theta cos^2\phi, \omega_2 = sin^2\theta sin^2\phi, \omega_3 = cos^2\theta$

The components of vector $\vec{u}$ can also be similarly written using angular variables as

$$u_1 = \sqrt{\tfrac{1}{3} - 2(\tfrac{1}{4}sin^2 2\theta sin^2\phi - r_1^2)}, u_2 = \sqrt{\tfrac{1}{3} - 2(\tfrac{1}{4}sin^2 2\theta cos^2\phi - r_2^2)}, u_3 = \sqrt{\tfrac{1}{3} - 2(\tfrac{1}{4}sin^4\theta sin^2 2\phi - r_3^2)}$$

### *Measure of Mixing*

The volume of the parallelepiped constructed from the 3 vectors is given by

$$V = |(\vec{u} \times \vec{v}) \cdot \vec{w}| = \tfrac{1}{\sqrt{3}} |\vec{u}| |\vec{v}| \sin \alpha$$

|

In Cartesian coordinates it is

$$V = \frac{1}{\sqrt{3}}\{(u_2 v_3 - u_3 v_2) + (u_3 v_1 - u_1 v_3) + (u_1 v_2 - u_2 v_1)\}$$

For pure state vectors $\vec{u}$ and $\vec{w}$ are of unit length and so $V$ becomes zero. So nonzero $V$ is a measure of mixing of a given qutrit state. For a given pair of vectors $(\vec{u}, \vec{v})$ the maximum mixed state occurs when $\alpha = \pi/2$.

## 4- The 2-Sections of the Qutrit State Space

In the 8-dimensional QtSS, visualization of state space is very difficult except in 2 and 3 dimensions. In reference [2, 3] those slices were derived and shown to have interesting geometrical shapes. Any attempts to go beyond the 3-dimensions proved almost impossible. The new vector representation is free of this problem and sections of all possible dimensions can be understood as affecting the three vectors in a specific manner. As an example, we will present the 2-sections in their terms to illustrate this point.

The number of 2-sections, when any of the two parameters are non-zero, is $8_{C_2}= 28$. They are distributed in 4 classes describing circle, triangle, parabola and ellipse shapes in the 8-dimensional QtSS.

We note that in the following analysis we always have, $w^2 = 1$ by definition.

(1) **Circle** (17 states): {12}, {13}, {23}, {14}, {15}, {16}, {17}, {24}, {25}, {26}, {27}, {45}, {46}, {47}, {56}, {57}, and {67}.

Example: $\{n_1 = 1, n_2 = 2\}$ *2-section in SU(3) and Spin-1 representation*

The equation for $I_2$ in this case in terms of SU(3) parameters is found as

SU (3) Representation: $n_1^2 + n_2^2 \leq 1$.

Spin-1 Representation: $q_3^2 + a_3^2 \leq \left(2/\sqrt{3}\right)^2$

This is the equation for a circle of radius $1/\sqrt{3}$ in SU (3) and 2/3 in Spin-1 representation. Similarly the $I_3$ equation gives

SU (3) Representation: $n_1^2 + n_2^2 = \left(1/\sqrt{3}\right)^2$.

Spin-1 Representation: $q_3^2 + a_3^2 = (2/3)^2$

The 3-dimensional vector components take the following forms

$$u_1 = u_2 = \tfrac{1}{3}, u_3 = \sqrt{\tfrac{1}{9} + \tfrac{1}{2}(q_3^2 + a_3^2)}, u^2 = \tfrac{1}{3} + \tfrac{1}{2}(q_3^2 + a_3^2) \leq 1,$$

$$v_1 = v_2 = \sqrt{\tfrac{7}{27}}, v_3 = \sqrt{\tfrac{7}{27} + \tfrac{1}{2}(q_3^2 + a_3^2)}, v^2 = \tfrac{7}{9} + \tfrac{1}{2}(q_3^2 + a_3^2) = 1,$$

To determine the purity of the states, we solve $v^2 = 1$ and substitute the result in $u^2$. This leads to $u^2 = \tfrac{5}{9} < 1$. So the circle 2-sections have no pure states.

So the $\{n_1 = 1, n_2 = 2\}$ section for a circle translates into constraints on the lengths and components (or equivalently on angles) of $\vec{u}$, and $\vec{v}$ vectors.

(2) **Triangle** (3 states): {18}, {28}, and {38}

Example: {18}

(i) $\{n_1, n_8\}$, SU(3) Representation Constraints

$n_1^2 + n_8^2 \leq 1, (1 + n_8 + \sqrt{3}n_3)(1 + n_8 - \sqrt{3}n_3)(1 - 2n_8) = 0$ at QtSS boundary

(ii) $\{\omega_1, q_3\}$, Spin-1 Representation Constraints:

$u_1 = u_2 = \omega_1, u_3 = \sqrt{(1 - 2\omega_1)^2 + \frac{1}{2}q_3^2}$,

$u^2 = 1 - 6\omega_1\left(\frac{2}{3} - \omega_1\right) + \frac{1}{2}q_3^2 \leq 1$

$v_1 = v_2 = \sqrt{3\omega_1^2 - 2\omega_1^3}, v_3 = \sqrt{(1 - 2\omega_1)\left\{(1 - 2\omega_1)(1 + 4\omega_1) + \frac{3}{2}q_3^2\right\}}$,

$v^2 = 1 - 12\left(\frac{1}{2} - \omega_1\right)\left(\omega_1^2 - \frac{1}{4}q_3^2\right) = 1$

**Pure States:** $v^2 = 1$ gives $\{\omega_1, q_3\} = \{0,0\}, \{\frac{1}{2}, 1\}$, and $\{\frac{1}{2}, -1\}$. Substitution gives $u^2 = 1$.
So all of these 3 states are pure.

(3) **Parabola** (4 states) {34}, {35}, {36}, and {37}

Example: {34}

    (i)    $\{n_3, n_4\}$, SU(3) Representation Constraints

$n_3^2 + n_4^2 \leq 1, (1 + \sqrt{3}n_3 - 3n_4^2)(1 - \sqrt{3}n_3) = 0$, equation for parabola

    (ii)    $\{\omega_1, q_2\}$, Spin-1 Representation Constraints:

$u_1 = \sqrt{\frac{2}{3}\omega_1 - \frac{1}{9}}, u_2 = \sqrt{\frac{1}{3} - \frac{2}{3}\omega_1 + \frac{1}{2}q_2^2}, u_3 = \sqrt{\frac{1}{3} - 2\omega_1(\frac{2}{3} - \omega_1)}$,

$u^2 = \frac{5}{9} + 2\omega_1\left(\omega_1 - \frac{2}{3}\right) + \frac{1}{2}q_2^2 \leq 1$,

$v_1 = v_3 = \sqrt{\frac{1}{3} - \frac{2}{3}\omega_1(\frac{2}{3} - \omega_1)}, v_2 = \sqrt{\frac{1}{3} - (\frac{2}{3} - \omega_1)(\frac{2}{3}\omega_1 - \frac{3}{2}q_2^2)}$,

$v^2 = 1 - \frac{1}{2}\left(\frac{2}{3} - \omega_1\right)(4\omega_1 - 3q_2^2) = 1$

**Pure States:** $v^2 = 1$ gives $\{\omega_1, q_2\} = \{\frac{2}{3}, \frac{2}{3}\sqrt{2}\}$ and $\{\frac{2}{3}, -\frac{2}{3}\sqrt{2}\}$
Substitution gives $u^2 = 1$. So these two states are pure.

(4) **Ellipse** (4 states): {48}, {58}, {68}, and {78}

Example: {48}

    (i)    SU(3) Representation Constraints

$n_4^2 + n_8^2 \leq 1, \left[3n_4^2 + 2\left(n_8 + \frac{1}{4}\right)^2 - \frac{9}{8}\right](1 + n_8) = 0$. Equation of ellipse

    (ii)    $\{\omega_1, q_2\}$, Spin-1 Representation Constraints:

$u_1 = \sqrt{\frac{1}{3} - \frac{2}{3}\omega_1(1-2\omega_1)}, u_2 = \sqrt{\frac{1}{3} - \frac{2}{3}\omega_1(1-2\omega_1) + \frac{1}{2}q_2^2}, u_3 = \sqrt{\frac{1}{3} - 2\omega_1^2}$,

$u^2 = 1 - 2\omega_1(2 - 3\omega_1) + \frac{1}{2}q_2^2 \leq 1$,

$v_1 = v_3 = \sqrt{\frac{1}{3} - 2\omega_1^2(1-2\omega_1)}, v_2 = \sqrt{\frac{1}{3} - 2\omega_1^2(1-2\omega_1) + 6\omega_1 q_2^2}$,

$v^2 = 1 - 6\omega_1^2(1-2\omega_1) + \frac{3}{2}\omega_1 q_2^2 = 1$,

**Pure States:** $v^2 = 1$ gives $\{\omega_1, q_2\} = \{0,0\}$, $\{\frac{1}{4}, \frac{1}{\sqrt{2}}\}$, and $\{\frac{1}{2}, 0\}$.
Substitution gives $u^2 = 1$. as pure states only for the first set

## 5- The 3-Sections of the Qutrit State Space

The following table gives similar information for 3-sections which includes 56 states.

(1) **Cone** (7 states): {128}, {138}, {238}, {348}, {358}, {368}, and {378}

Example: {128}

    (i)     SU(3) Representation Constraints

$n_1^2 + n_2^2 + n_8^2 \leq 1$, $[(n_8 + 1)^2 - 3(n_1^2 + n_2^2)](2n_8 - 1) = 0$.

Eqn. of cone truncated by a plane

    (ii)     $\{\omega_1, q_3, a_3\}$, Spin-1 Representation Constraints:

$$u_1 = \sqrt{\frac{1}{3} - \frac{2}{3}\omega_1(1-2\omega_1)} = u_2, u_3 = \sqrt{\frac{1}{3} - 2\omega_1^2 + \frac{1}{2}(q_3^2 + a_3^2)},$$

$$u^2 = 1 - 6\omega_1\left(\frac{2}{3} - \omega_1\right) + \frac{1}{2}(q_3^2 + a_3^2) \leq 1$$

$$v_1 = v_2 = \sqrt{\frac{1}{3} - 2\omega_1^2(1-2\omega_1)}, v_2 = \sqrt{\frac{1}{3} + (1-2\omega_1)\{-2\omega_1^2 + \frac{3}{2}(q_3^2 + a_3^2)\}},$$

$$v^2 = 1 - 12\left(\frac{1}{2} - \omega_1\right)\left\{\omega_1^2 - \frac{1}{4}(q_3^2 + a_3^2)\right\} = 1$$

**Pure States:** $v^2 = 1$ gives $\{\omega_1, q_3, a_3\} = \{0,0,0\}$, and $\{\omega_1, \sqrt{q_3^2 + a_3^2}\} = \{\frac{1}{2}, 1\}$ and both give solutions satisfying $u^2 = 1$. So these are pure states with the last solution giving pure states on a circle.

(2) **Paraboloid** (2 states): {345} and {367}

Example: {345}

    (i)     SU(3) Representation Constraints

$n_3^2 + n_4^2 + n_5^2 \leq 1$, $[3(n_4^2 + n_5^2) - (\sqrt{3}n_3 + 1)](1 - \sqrt{3}n_3) = 0$.

Eqn. of paraboloid truncated by a plane

    (ii)     $\{\omega_1, q_2, a_2\}$, Spin-1 Representation Constraints:

$$u_1 = \sqrt{\frac{1}{3} - \frac{2}{3}\left(\frac{2}{3} - \omega_1\right)}, u_2 = \sqrt{\frac{1}{3} - \frac{2}{3}\omega_1 + \frac{1}{2}(q_2^2 + a_2^2)}, u_3 = \sqrt{\frac{1}{3} - 2\omega_1\left(\frac{2}{3} - \omega_1\right)}$$

$$u^2 = \frac{5}{9} - 2\omega_1\left(\frac{2}{3} - \omega_1\right) + \frac{1}{2}(q_2^2 + a_2^2) \leq 1$$

$$v_1 = v_3 = \sqrt{\frac{1}{3} - \frac{2}{3}\omega_1\left(\frac{2}{3} - \omega_1\right)}, v_2 = \sqrt{\frac{1}{3} + \left(\frac{2}{3} - \omega_1\right)\{-\frac{2}{3}\omega_1 + \frac{3}{2}(q_2^2 + a_2^2)\}},$$

$$v^2 = 1 + 2\left(\frac{2}{3} - \omega_1\right)\left\{\omega_1 - \frac{3}{4}(q_2^2 + a_2^2)\right\} = 1$$

**Pure States:** $v^2 = 1$ gives $\{\omega_1, \sqrt{q_2^2 + a_2^2}\} = \{\frac{2}{3}, \frac{2\sqrt{2}}{3}\}$. It satisfies $u^2 = 1$, leading to a set of pure states on a circle.

(3) **Ellipsoid** (6 states) {458}, {468}, {478}, {568}, {578}, and {678}

Example: {468}

    (i)    SU(3) Representation Constraints

$n_4^2 + n_6^2 + n_8^2 \leq 1$, $\left[3(n_4^2 + n_6^2) + 2\left(n_8 + \frac{1}{4}\right)^2 - \frac{9}{8}\right](1 + n_8) = 0$.

Eqn. of ellipsoid of revolution

    (ii)    $\{\omega_1, q_1, q_2\}$, Spin-1 Representation Constraints:

$u_1 = \sqrt{\frac{1}{3} - 2\omega_1(1-2\omega_1) + \frac{1}{2}q_1^2}$, $u_2 = \sqrt{\frac{1}{3} - 2\omega_1(1-2\omega_1) + \frac{1}{2}q_2^2}$, $u_3 = \sqrt{\frac{1}{3} - 2\omega_1^2}$

$u^2 = 1 - 6\omega_1\left(\frac{2}{3} - \omega_1\right) + \frac{1}{2}(q_1^2 + q_2^2) \leq 1$

$v_1 = \sqrt{\frac{1}{3} - 2\omega_1^2(1-2\omega_1) + \frac{3}{2}\omega_1 q_1^2}$, $v_2 = \sqrt{\frac{1}{3} - 2\omega_1^2(1-2\omega_1) + \frac{3}{2}\omega_1 q_2^2}$, $v_3 = \sqrt{\frac{1}{3} - 2\omega_1^2(1-2\omega_1)}$

$v^2 = 1 - 12\omega_1^2\left(\frac{1}{2} - \omega_1\right) + \frac{3}{2}\omega_1(q_1^2 + q_2^2) = 1$

**Pure States:** $v^2 = 1$ gives $\{\omega_1, q_1, q_2\} = \{0,0,0\}$ satisfying $u^2 = 1$. This gives one pure state.

(4) **Obese Tetrahedron** (8 states): {146}, {157}, {247}, {256}, {346}, {347}, {356}, and {357}

Example: {146}

    (i)    SU(3) Representation Constraints

$n_1^2 + n_4^2 + n_6^2 \leq 1$, $\left[3(n_1^2 + n_4^2 + n_6^2) - 6\sqrt{3}n_1 n_4 n_6\right] = 1$.

Eqn. of obese-tetrahedron

    (ii)    $\{q_1, q_2, q_3\}$, Spin-1 Representation Constraints:

$u_1 = \sqrt{\frac{1}{9} + \frac{1}{2}q_1^2}$, $u_2 = \sqrt{\frac{1}{9} + \frac{1}{2}q_2^2}$, $u_3 = \sqrt{\frac{1}{9} + \frac{1}{2}q_3^2}$

$u^2 = \frac{1}{3} + \frac{1}{2}(q_1^2 + q_2^2 + q_3^2) \leq 1$

$v_1 = \sqrt{\frac{7}{27} + \frac{1}{2}q_1 q_2 q_3 + \frac{1}{2}q_1^2}$, $v_2 = \sqrt{\frac{7}{27} + \frac{1}{2}q_1 q_2 q_3 + \frac{1}{2}q_2^2}$, $v_3 = \sqrt{\frac{7}{27} + \frac{1}{2}q_1 q_2 q_3 + \frac{1}{2}q_3^2}$

$v^2 = \frac{7}{9} + \frac{1}{2}(q_1^2 + q_2^2 + q_3^2) + \frac{3}{2}q_1 q_2 q_3 = 1$

**Pure States:** $v^2 = 1$ leads to 4 pure states given by $\{q_1, q_2, q_3\} = \frac{2}{3}\{-1,1,1\}$, $\frac{2}{3}\{1,-1,1\}$, $\frac{2}{3}\{1,1,-1\}$, and $\frac{2}{3}\{-1,-1,-1\}$, all satisfying $u^2 = 1$

(5) **RS1** (8 states): {134}, {135}, {136}, {137}, {234}, {235}, {236}, and {237}

Example: {134}

    (i)    SU(3) Representation Constraints

$n_1^2 + n_3^2 + n_4^2 \leq 1$, $\left[3(n_1^2 + n_3^2 + n_4^2) - 3\sqrt{3}n_4^2 n_3\right] = 1$.

Eqn. of obese-tetrahedron

(ii) $\{\omega_1, q_2, q_3\}$, Spin-1 Representation Constraints:

$u_1 = \sqrt{\frac{1}{3} - \frac{2}{3}\left(\frac{2}{3} - 2\omega_1\right)}, u_2 = \sqrt{\frac{1}{3} - \frac{2}{3}\omega_1 + \frac{1}{2}q_2^2}, u_3 = \sqrt{\frac{1}{3} - 2\omega_1\left(\frac{2}{3} - \omega_1\right) + \frac{1}{2}q_3^2}$

$u^2 = 1 + 2\left(\omega_1^2 - \frac{2}{3}\omega_1 - \frac{2}{9}\right) + \frac{1}{2}(q_2^2 + q_3^2) \leq 1$

$v_1 = \sqrt{\frac{1}{3} - \frac{2}{3}\omega_1\left(\frac{2}{3} - 2\omega_1\right)}, v_2 = \sqrt{\frac{1}{3} - \omega_1\left(\frac{2}{3} - 2\omega_1\right)\left(\frac{2}{3} - \frac{3}{2}q_2^2\right)}, v_3 = \sqrt{\frac{1}{3} - \frac{2}{3}\omega_1\left(\frac{2}{3} - 2\omega_1\right) + \frac{1}{2}q_3^2}$

$v^2 = 1 - 6\left(\frac{2}{3} - \omega_1\right)\left(\frac{\omega_1}{3} - \frac{1}{4}q_2^2\right) + \frac{1}{2}q_3^2 = 1$

**Pure States:** $v^2 = 1$ gives $\{\omega_1, q_2, q_3\} = \{\frac{2}{3}, \pm\frac{2\sqrt{2}}{3}, 0\}$ satisfying $u^2 = 1$. This gives two pure states.

(6) **RS2** (8 states): {148}, {158}, {168}, {178}, {248}, {258}, {268}, and {278}

Example: {148}

(i) SU(3) Representation Constraints

$n_1^2 + n_4^2 + n_8^2 \leq 1, \left[3(n_1^2 + n_4^2 + n_8^2) - 3n_8(2n_1^2 - n_4^2 - \frac{2}{3}n_8^2)\right] = 1$.

Eqn. of obese-tetrahedron

(ii) $\{\omega_1, q_2, q_3\}$, Spin-1 Representation Constraints:

$u_1 = \sqrt{\frac{1}{3} - 2\omega_1(1 - 2\omega_1)}, u_2 = \sqrt{\frac{1}{3} - 2\omega_1(1 - 2\omega_1) + \frac{1}{2}q_2^2}, u_3 = \sqrt{\frac{1}{3} - 2\omega_1^2 + \frac{1}{2}q_3^2}$

$u^2 = 1 - 2\omega_1(2 - 3\omega_1) + \frac{1}{2}(q_2^2 + q_3^2) \leq 1$

$v_1 = \sqrt{\frac{1}{3} - 2\omega_1^2(1 - 2\omega_1)}, v_2 = \sqrt{\frac{1}{3} - 2\omega_1^2(1 - 2\omega_1) + \frac{3}{2}\omega_1 q_2^2}, v_3 = \sqrt{\frac{1}{3} + (1 - 2\omega_1)(-2\omega_1^2 + \frac{3}{2}q_2^2)}$

$v^2 = 1 - (1 - 2\omega_1)\left(6\omega_1^2 - \frac{3}{2}q_3^2\right) + \frac{3}{2}\omega_1 q_1^2 = 1$

**Pure States:** $v^2 = 1$ gives $\{\omega_1, q_2, q_3\} = \{0,0,0\}$ and $\{\frac{1}{2}, 0, \pm 1\}$ both satisfying $u^2 = 1$. This gives two pure states.

(7) **Sphere** (17 states): {123}, {124}, {125}, {126}, {127}, {145}, {147}, {156}, {167}, {245}, {246}, {257}, {267}, {456}, {457}, {467}, and {567}

Example: {123}

(i) SU(3) Representation Constraints

$n_1^2 + n_2^2 + n_3^2 \leq 1, 3(n_1^2 + n_2^2 + n_3^2) = 1$. Eqn. of a sphere

(ii) $\{\omega_1, a_3, q_3\}$, Spin-1 Representation Constraints:

$u_1 = \sqrt{-\frac{1}{9} + \frac{2}{3}\omega_1}, u_2 = \sqrt{\frac{1}{3} - \frac{2}{3}\omega_1}, u_3 = \sqrt{\frac{1}{3} - 2\omega_1(\frac{2}{3} - \omega_1) + \frac{1}{2}(q_3^2 + a_3^2)}$

$u^2 = \frac{5}{9} - \frac{4}{3}\omega_1 + 2\omega_1^2 + \frac{1}{2}(q_3^2 + a_3^2) \leq 1$

$v_1 = v_2 = \sqrt{\frac{1}{3} - \frac{2}{3}\omega_1(\frac{2}{3} - \omega_1)}, v_3 = \sqrt{\frac{1}{3} - \frac{2}{3}\omega_1\left(\frac{2}{3} - \omega_1\right) + \frac{1}{2}(q_3^2 + a_3^2)}$

$v^2 = 1 - 2\omega_1\left(\frac{2}{3} - \omega_1\right) + \frac{1}{2}(q_3^2 + a_3^2) = 1$

**Pure States:** There are no pure states in this type.

# 6-Higher order-Sections

Real power of these 3-dimensional vectors is seen when we need to analyze the behavior of higher order sections. In general, they lead to shapes in dimensions higher than 3 and thus are almost impossible to visualize. Below we give one example each of the higher order sections in the table. This emphasizes the utility of this approach where the difficulty of visualizing them has been replaced with the ease of handling 3-dimensional vectors.

(1) **4-Sections** (70 states):

Example: {1234}:

(i) SU(3) Representation Constraints

$n_1^2 + n_2^2 + n_3^2 + n_4^2 \leq 1, 3(n_1^2 + n_2^2 + n_3^2) + 3n_4^2\left(1 - \sqrt{3}n_3\right) = 1.$

(ii) $\{\omega_1, q_2, q_3, a_3\}$, Spin-1 Representation Constraints:

Here $n_8 = 0, \omega_2 = \frac{2}{3} - \omega_1, \omega_3 = \frac{1}{3}, n_3 = \sqrt{3}(\omega_1 - \frac{1}{3})$

$u_1 = \sqrt{\frac{2}{3}\omega_1 - \frac{1}{9}}, u_2 = \sqrt{\frac{1}{3} - \frac{2}{3}\omega_1 + \frac{1}{2}q_2^2}, u_3 = \sqrt{\frac{1}{3} - 2\omega_1\left(\frac{2}{3} - \omega_1\right) + \frac{1}{2}(q_3^2 + a_3^2)},$

$u^2 = \frac{5}{9} - 2\omega_1\left(\frac{2}{3} - \omega_1\right) + \frac{1}{2}(q_2^2 + q_3^2 + a_3^2) \leq 1$

$v_1 = \sqrt{\frac{1}{3} - \frac{2}{3}\omega_1\left(\frac{2}{3} - \omega_1\right)}, v_2 = \sqrt{\frac{1}{3} + \left(\frac{2}{3} - \omega_1\right)\{-\frac{2}{3}\omega_1 + \frac{3}{2}q_2^2\}}, v_3 = \sqrt{\frac{1}{3} - \frac{2}{3}\omega_1\left(\frac{2}{3} - \omega_1\right) + \frac{1}{2}(q_3^2 + a_3^2)}$

$v^2 = 1 - 2\omega_1\left(\frac{2}{3} - \omega_1\right) - \frac{3}{2}\omega_1 q_2^2 + \frac{1}{2}(2q_2^2 + q_3^2 + a_3^2) = 1$

**(2) 5-Section** (56 states)

Example: {12345}

(i) SU(3) Representation Constraints

$n_1^2 + n_2^2 + n_3^2 + n_4^2 + n_5^2 \leq 1, 3(n_1^2 + n_2^2 + n_3^2) + 3(n_4^2 + n_5^2)\left(1 - \sqrt{3}n_3\right) = 1.$

(ii) $\{\omega_1, q_2, a_2, q_3, a_3\}$, Spin-1 Representation Constraints:

Here $n_8 = 0, \omega_2 = \frac{2}{3} - \omega_1, \omega_3 = \frac{1}{3}, n_3 = \sqrt{3}(\omega_1 - \frac{1}{3})$

$u_1 = \sqrt{\frac{2}{3}\omega_1 - \frac{1}{9}}, u_2 = \sqrt{\frac{1}{3} - \frac{2}{3}\omega_1 + \frac{1}{2}(q_2^2 + a_2^2)}, u_3 = \sqrt{\frac{1}{3} - 2\omega_1\left(\frac{2}{3} - \omega_1\right) + \frac{1}{2}(q_3^2 + a_3^2)}$

$u^2 = \frac{5}{9} - 2\omega_1\left(\frac{2}{3} - \omega_1\right) + \frac{1}{2}(q_2^2 + a_2^2 + q_3^2 + a_3^2) \leq 1$

$v_1 = \sqrt{\frac{1}{3} - \frac{2}{3}\omega_1\left(\frac{2}{3} - \omega_1\right)}, v_2 = \sqrt{\frac{1}{3} + \left(\frac{2}{3} - \omega_1\right)\{-\frac{2}{3}\omega_1 + \frac{3}{2}(q_2^2 + a_2^2)\}}, v_3 = \sqrt{\frac{1}{3} - \frac{2}{3}\omega_1\left(\frac{2}{3} - \omega_1\right) + \frac{1}{2}(q_3^2 + a_3^2)}$

$v^2 = 1 - 2\omega_1\left(\frac{2}{3} - \omega_1\right) - \frac{3}{2}\omega_1(q_2^2 + a_2^2) + \frac{1}{2}(2q_2^2 + 2a_2^2 + q_3^2 + a_3^2) = 1$

**(3) 6-Section** (28 states)

Example: {123456}

(i) SU(3) Representation Constraints

$$n_1^2 + n_2^2 + n_3^2 + n_4^2 + n_5^2 + n_6^2 \leq 1,$$

$$3(n_1^2 + n_2^2 + n_3^2) + 3(n_4^2 + n_5^2)\left(1 - \sqrt{3}n_3\right) - 6\sqrt{3}n_6(n_1n_4 + n_2n_5) + 3n_6^2\left(1 + \sqrt{3}n_3\right) = 1$$

(ii) $\{\omega_1, q_1, q_2, a_2, q_3, a_3\}$, Spin-1 Representation Constraints:

Here $n_8 = 0, \omega_2 = \frac{2}{3} - \omega_1, \omega_3 = \frac{1}{3}, n_3 = \sqrt{3}(\omega_1 - \frac{1}{3})$

$$u_1 = \sqrt{\frac{2}{3}\omega_1 - \frac{1}{9} + \frac{1}{2}q_1^2},\ u_2 = \sqrt{\frac{1}{3} - \frac{2}{3}\omega_1 + \frac{1}{2}(q_2^2 + a_2^2)},\ u_3 = \sqrt{\frac{1}{3} - 2\omega_1\left(\frac{2}{3} - \omega_1\right) + \frac{1}{2}(q_3^2 + a_3^2)}$$

$$u^2 = \frac{5}{9} - 2\omega_1\left(\frac{2}{3} - \omega_1\right) + \frac{1}{2}(q_1^2 + q_2^2 + a_2^2 + q_3^2 + a_3^2) \leq 1$$

$$v_1 = \sqrt{\frac{1}{3} - \frac{2}{3}\omega_1\left(\frac{2}{3} - \omega_1\right) + \frac{1}{2}q_1(q_2q_3 - a_2a_3) + \frac{3}{2}\omega_1 q_1^2},$$

$$v_2 = \sqrt{\frac{1}{3} - \frac{2}{3}\omega_1\left(\frac{2}{3} - \omega_1\right) + \frac{1}{2}q_1(q_2q_3 - a_2a_3) + \frac{3}{2}\left(\frac{2}{3} - \omega_1\right)(q_2^2 + a_2^2)},$$

$$v_3 = \sqrt{\frac{1}{3} - \frac{2}{3}\omega_1\left(\frac{2}{3} - \omega_1\right) + \frac{1}{2}q_1(q_2q_3 - a_2a_3) + \frac{1}{2}(q_3^2 + a_3^2)}$$

$$v^2 = 1 - 2\omega_1\left(\frac{2}{3} - \omega_1\right) + \frac{3}{2}q_1(q_2q_3 - a_2a_3) - \frac{3}{2}\omega_1(q_2^2 + a_2^2 - q_1^2) + \frac{1}{2}(2q_2^2 + 2a_2^2 + q_3^2 + a_3^2) = 1$$

(4) **7-Section** (8 states)

Example: {1234567}

(i) SU(3) Representation Constraints

$$n_1^2 + n_2^2 + n_3^2 + n_4^2 + n_5^2 + n_6^2 + n_7^2 \leq 1, 3(n_1^2 + n_2^2 + n_3^2) + 3(n_4^2 + n_5^2)\left(1 - \sqrt{3}n_3\right) - 6\sqrt{3}\{n_6(n_1n_4 + n_2n_5) + n_7(n_1n_5 - n_2n_4)\} + 3(n_6^2 + n_7^2)\left(1 + \sqrt{3}n_3\right) = 1$$

(ii) $\{\omega_1, q_1, a_1, q_2, a_2, q_3, a_3\}$, Spin-1 Representation Constraints:

Here $n_8 = 0, \omega_2 = \frac{2}{3} - \omega_1, \omega_3 = \frac{1}{3}, n_3 = \sqrt{3}(\omega_1 - \frac{1}{3})$

$$u_1 = \sqrt{\frac{2}{3}\omega_1 - \frac{1}{9} + \frac{1}{2}(q_1^2 + a_1^2)},\ u_2 = \sqrt{\frac{1}{3} - \frac{2}{3}\omega_1 + \frac{1}{2}(q_2^2 + a_2^2)},\ u_3 = \sqrt{\frac{1}{3} - 2\omega_1\left(\frac{2}{3} - \omega_1\right) + \frac{1}{2}(q_3^2 + a_3^2)}$$

$$u^2 = \frac{5}{9} - 2\omega_1\left(\frac{2}{3} - \omega_1\right) + \frac{1}{2}(q_1^2 + a_1^2 + q_2^2 + a_2^2 + q_3^2 + a_3^2) \leq 1$$

$$v_1 = \sqrt{\frac{1}{3} - \frac{2}{3}\omega_1\left(\frac{2}{3} - \omega_1\right) + \frac{1}{2}q_1(q_2q_3 - a_2a_3) + \frac{3}{2}\omega_1 q_1^2},$$

$$v_2 = \sqrt{\frac{1}{3} - \frac{2}{3}\omega_1\left(\frac{2}{3} - \omega_1\right) + \frac{1}{2}q_1(q_2q_3 - a_2a_3) + \frac{3}{2}\left(\frac{2}{3} - \omega_1\right)(q_2^2 + a_2^2)},$$

$$v_3 = \sqrt{\frac{1}{3} - \frac{2}{3}\omega_1\left(\frac{2}{3} - \omega_1\right) + \frac{1}{2}q_1(q_2q_3 - a_2a_3) + \frac{1}{2}(q_3^2 + a_3^2)}$$

$$v^2 = 1 - 2\omega_1\left(\frac{2}{3} - \omega_1\right) + \frac{3}{2}q_1(q_2q_3 - a_2a_3) - \frac{3}{2}\omega_1(q_2^2 + a_2^2 - q_1^2) + \frac{1}{2}(2q_2^2 + 2a_2^2 + q_3^2 + a_3^2) = 1$$

**(5) 8-Section** (1 state)

Example: {12345678}: This is the full 8-dimensional space in the SU (3) or Spin-1 representation so starting vector definitions in full form apply.

(i) SU(3) Representation Constraints

$$I_2 = n_1^2 + n_2^2 + n_3^2 + n_4^2 + n_5^2 + n_6^2 + n_7^2 + n_8^2 \leq 1,$$

$$I_3 = 3\sum_{i=1}^{8} n_i^2 - 6n_8\left(\sum_{i=1}^{3} n_i^2 - \frac{1}{2}\sum_{i=4}^{7} n_i^2 - \frac{1}{3}n_8^2\right) - 6\sqrt{3}(n_1 n_4 n_6 + n_1 n_5 n_7 + n_2 n_5 n_6 - n_2 n_4 n_7)$$
$$- 3\sqrt{3}n_3(n_4^2 + n_5^2 - n_6^2 - n_7^2) = 1$$

(ii) $\{\omega_1, \omega_2, q_1, a_1, q_2, a_2, q_3, a_3\}$, Spin-1 Representation Constraints:

Here $\omega_3 = 1 - (\omega_1 + \omega_2), n_3 = \frac{\sqrt{3}}{2}(\omega_1 - \omega_2), n_8 = \frac{3}{2}(\omega_1 + \omega_2) - 1$.

$$I_2 = \sum_{i=1}^{3}(\omega_i^2 + 2r_i^2) \leq 1$$

$$u_1 = \sqrt{\frac{1}{3} - 2(\omega_2\omega_3 - r_1^2)}, u_2 = \sqrt{\frac{1}{3} - 2(\omega_3\omega_1 - r_2^2)}, u_3 = \sqrt{\frac{1}{3} - 2(\omega_1\omega_2 - r_3^2)}$$

Define: $D \equiv a_2 a_3 q_1 + a_3 a_1 q_2 + a_1 a_2 q_3 - q_1 q_2 q_3$. and $X \equiv 1/3 - 2\omega_1\omega_2\omega_3 - \frac{1}{2}D$. Then

$$I_3 = 1 - 6\omega_1\omega_2\omega_3 + 6\sum_{i=1}^{3}\omega_i r_i^2 - \frac{3}{2}D = 1,$$

$$v_1 = \sqrt{X + 6\omega_1 r_1^2}, v_2 = \sqrt{X + 6\omega_2 r_2^2}, v_3 = \sqrt{X + 6\omega_3 r_3^2}$$

Similarly higher-order sections lead to corresponding expressions for $\vec{u}$, $\vec{v}$, and $\vec{w}$-vector lengths and individual components. This will lead to a catalog of all the higher-order sections in terms of these vector expressions.

## 7-Conclusions

In this work we have shown that the 8-dimensional QtSS can be alternatively visualized in 3 dimensions using 3-dimensional vectors. It is seen that 2 and other higher order QtSS sections can be expressed as constraints on the directions and lengths of these vectors. Implications of this approach will be extended to the Heisenberg-Weyl (HW) and Mutually Unbiased Bases (MUB) representations for 1-qutrit QtSS and 2-qutrit entangled states in future.